\documentclass[%
reprint,
superscriptaddress,
 amsmath,amssymb,
 aps,
 prl,
]{revtex4-1}
\usepackage{dcolumn}
\usepackage{bm}
\usepackage{graphicx}
\usepackage{color}
\usepackage{comment}

\newcommand{\blue}{\textcolor{blue}}

\begin{document}

\preprint{}

\title{Distribution of the time of the maximum for stationary processes}
\author{Francesco Mori }
\affiliation{LPTMS, CNRS, Univ. Paris-Sud, Universit\'e Paris-Saclay, 91405 Orsay, France}
\author{Satya N. Majumdar}
\affiliation{LPTMS, CNRS, Univ. Paris-Sud, Universit\'e Paris-Saclay, 91405 Orsay, France}
\author{Gr\'egory Schehr }
\affiliation{Sorbonne Universit\'e, Laboratoire de Physique Th\'eorique et Hautes Energies, CNRS, UMR 7589, 4 Place Jussieu, 75252 Paris Cedex 05, France}

\date{\today}


\begin{abstract}
We consider a one-dimensional stationary stochastic process $x(\tau)$ of duration $T$. We study the probability density function (PDF) $P(t_{\rm m}|T)$ of the time $t_{\rm m}$ at which $x(\tau)$ reaches its global maximum. By using a path integral method, we compute $P(t_{\rm m}|T)$ for a number of equilibrium and nonequilibrium  stationary processes, including the Ornstein-Uhlenbeck process, Brownian motion with stochastic resetting and a single confined run-and-tumble particle. For a large class of equilibrium stationary processes that correspond to diffusion in a confining potential, we show that the scaled distribution $P(t_{\rm m}|T)$, for large $T$, has a universal form (independent of the details of the potential). This universal distribution is uniform in the ``bulk'', i.e., for $0 \ll t_{\rm m} \ll T$  and has a nontrivial edge scaling behavior for $t_{\rm m} \to 0$ (and when $t_{\rm m} \to T$), that we compute exactly. Moreover, we show that for any equilibrium process the PDF $P(t_{\rm m}|T)$ is symmetric around $t_{\rm m}=T/2$, i.e., $P(t_{\rm m}|T)=P(T-t_{\rm m}|T)$. This symmetry provides a simple method to decide whether a given stationary time series $x(\tau)$ is at equilibrium or not.
\end{abstract}


\maketitle

\newpage

\begin{figure}[t]
\includegraphics[scale=0.55]{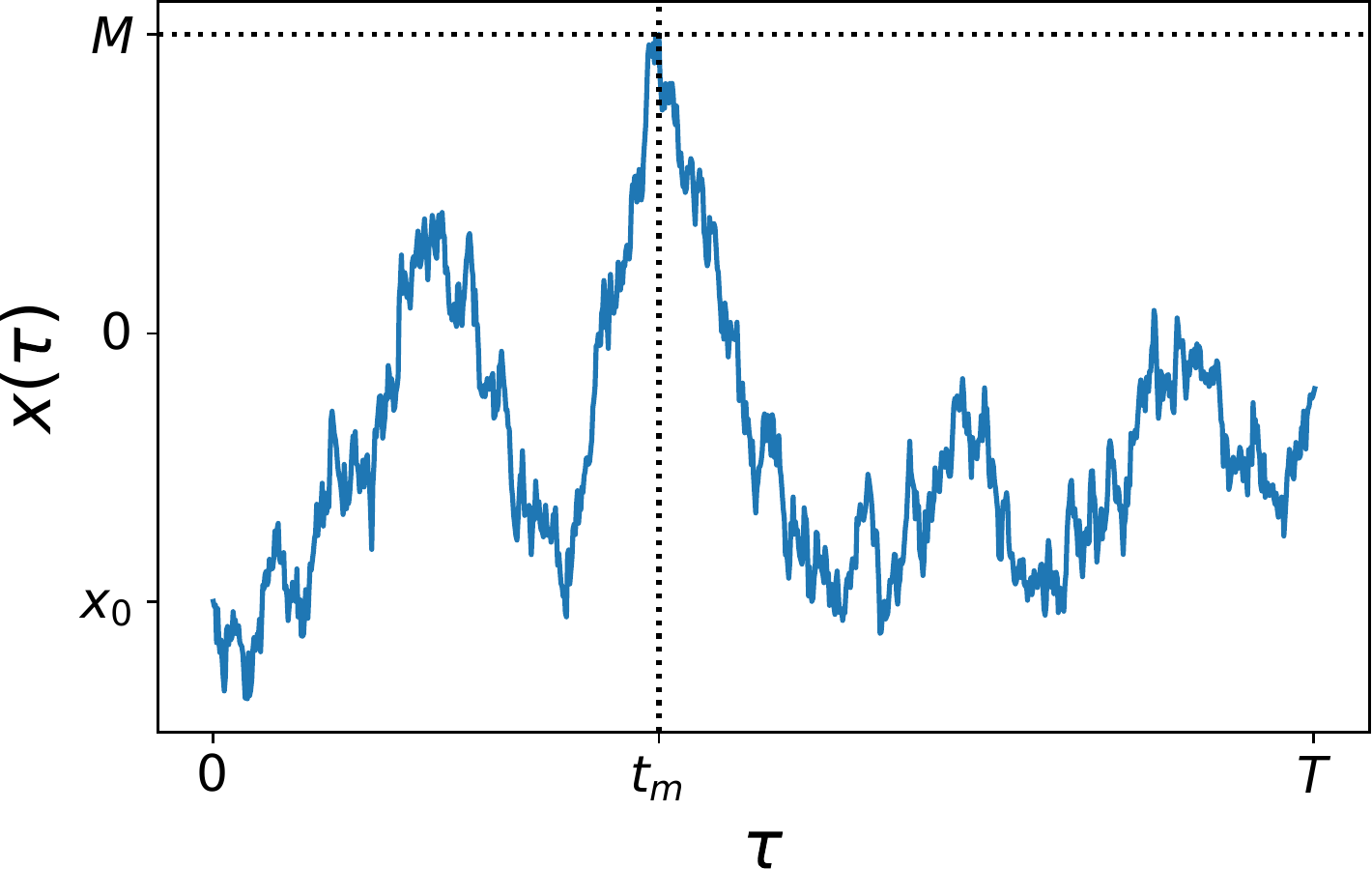} 
\caption{\label{fig:trajectory} Typical trajectory of a stationary process $x(\tau)$ as a function of time $\tau$. The process starts from position $x_0$, drawn from the stationary state, at the initial time and reaches the maximum $M$ at time $t_{\rm m}$.}
\end{figure}

The properties of extremes of stochastic processes are of fundamental importance in a wide range of practical situation, including finance, computer science, and climate science \cite{MP20}. For instance, in the context of climate change, it is paramount to estimate the probability of extreme climate events, such as heat waves, hurricanes, and tsunamis. Even if in many cases one is interested in the magnitude $M$ of such anomalous events, it is often  also relevant to study the time $t_{\rm m}$ at which they occur within some fixed time period $[0,T]$ (see Fig. \ref{fig:trajectory}). This observable $t_{\rm m}$, the time of the maximum, is a central quantity in several applications \cite{MP20,DW80,BC04,MRZ10,CKR15,BRM18,Dey21}. For instance, in finance, the distribution of the time at which the price of a stock attains its maximal value within a fixed time period is a quantity of clear practical interest.

Within the framework of extreme value theory, the statistical properties of $t_{\rm m}$ have been investigated for a wide range of stochastic processes \cite{levy40,feller50,SA53,She79,B03,LDM03,RFM07,MB08,MRK08,SLD10,MY10,M10,MRZ10a,RMC09,MCR10,RS11,DMRZ13,DW16,SDW18,SW21,SK19,MMS19,MLD20,MMS20,LM20,MLD20a}. For instance, in the case of an overdamped Brownian particle in one dimension, the full probability density function (PDF) $P(t_{\rm m}|T)$ of $t_{\rm m}$ was computed analytically by L\'evy and is given by \cite{levy40,feller50,SA53}
\begin{equation}
P(t_{\rm m}|T)=\frac{1}{\pi\sqrt{t_{\rm m}(T-t_{\rm m})}}\,,
\end{equation}
with $0\leq t_{\rm m}\leq T$. More recently, the PDF of $t_{\rm m}$ was also computed for constrained Brownian motions (BM) \cite{She79,B03,RFM07,MB08,MRK08,SLD10,MY10,MMS19,MMS20}, Bessel process \cite{SLD10}, and run-and-tumble particles (RTPs) \cite{SK19,MLD20a,MLD20}, amongst others. However, to the best of our knowledge, the time of the maximum has never been studied for stationary processes, i.e., stochastic processes whose statistical properties are invariant under a time shift.

Stationary phenomena are ubiquitous in nature and appear in a wide range of systems, including Brownian engines \cite{D97}, active matter \cite{C12} and climate systems \cite{WFM20}. They are divided into two main categories: equilibrium and out-of-equilibrium. Equilibrium processes satisfy the detailed balance condition, implying that all currents vanish and that the dynamics is time-reversible. Standard techniques from statistical physics can be applied to study equilibrium systems and thus their behavior is generally well understood. In contrast, nonequilibrium phenomena are characterized by the presence of currents in the steady state and very few general results exist in this case~\cite{jarzynski,K98,C1999,seifert05,seifert12,HG20}.

The distribution of the maximal value $M$ has been studied for several stationary processes of fixed duration $T$, including fluctuating interfaces~\cite{MC04,MC05}, the Ornstein-Uhlenbeck process \cite{MP20}, and BM with stochastic resetting \cite{EM11,MP20,MMSS21}. BM with stochastic resetting has become a rather popular subject of late both theoretically and experimentally -- for a recent review see \cite{EMS20}. Notably, in the case where the autocorrelation function of the process decays sufficiently fast the distribution of $M$, properly centered and scaled, approaches a universal Gumbel form at late times, i.e., for $T\gg T_{\rm micro}$ where $T_{\rm micro}$ is a microscopic correlation time \cite{Berman64,MC05,MP20}. However, it is not clear if this universality also extends to the distribution of the time $t_{\rm m}$ at which the maximum $M$ is reached. Moreover, since the statistical properties of stationary processes by definition do not evolve in time, one could naively expect the distribution of $t_{\rm m}$ to be uniform, i.e., $P(t_{\rm m}|T)=1/T$. Quite surprisingly, we show that this is not true in general, due to the presence of nonzero temporal correlations of the process.

\begin{figure}[t]
\includegraphics[scale=0.29]{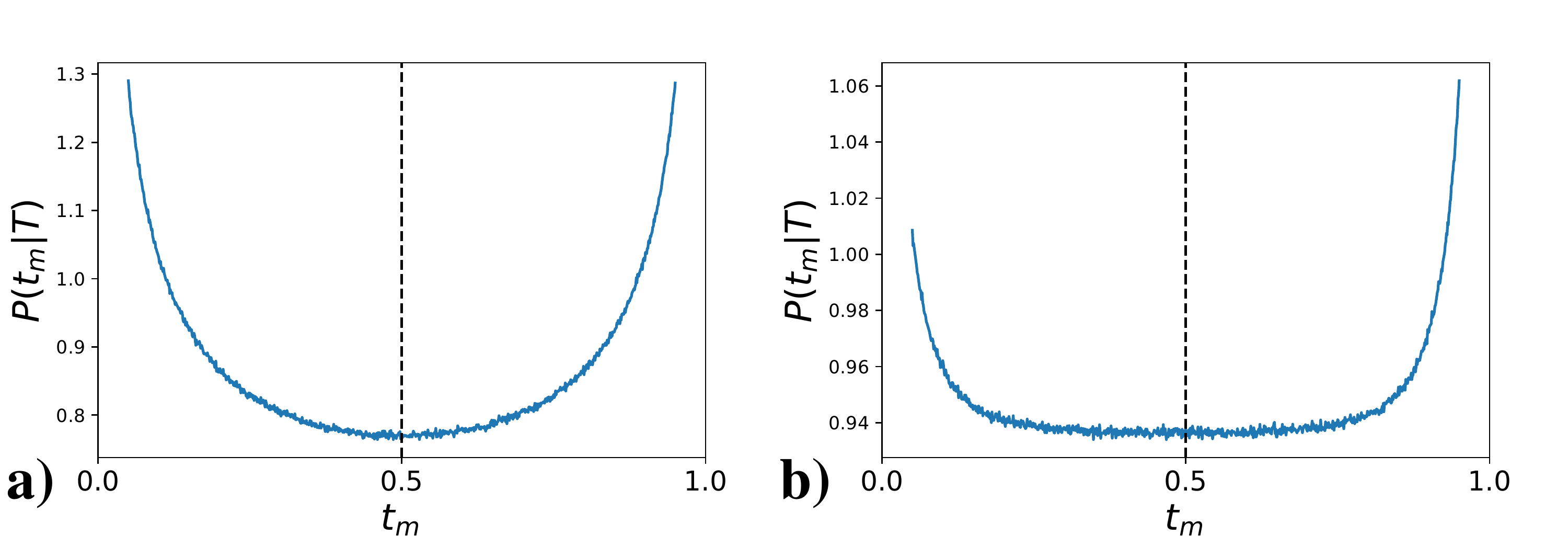} 
\caption{\label{fig:small_t} {\bf a)} Probability distribution $P(t_{\rm m}|T)$ versus $t_{\rm m}$ in the case of the Ornstein-Uhlenbeck process, obtained from numerical simulations with $\alpha=D=T=1$. The PDF $P(t_{\rm m}|T)$ is symmetric around the middle point $t_{\rm m}=T/2$ (dashed black line), since the process is at equilibrium.  {\bf b)} Probability distribution $P(t_{\rm m}|T)$ versus $t_{\rm m}$ in the case of Brownian motion with resetting, obtained from numerical simulations with $D=T=1$ and $r=10$. The PDF $P(t_m|T)$ is asymmetric around the middle point $t_{\rm m}=T/2$ (for further proof, see \cite{supmat}), signaling the nonequilibrium nature of the process.}
\end{figure}

In this Letter, we consider a one-dimensional stationary stochastic process $x(\tau)$, evolving in the time interval $[0,T]$ (see Fig. \ref{fig:trajectory}). At the initial time, we assume that the process has already reached its stationary state $P_{\rm st}(x)$. This is equivalent to preparing the system in some initial condition at time $\tau=-\infty$ and starting to observe it at $\tau=0$. Using a path-integral approach, we compute exactly the distribution $P(t_{\rm m}|T)$ of the time $t_{\rm m}$ at which $x(\tau)$ attains its maximal value for several stationary models, both equilibrium and nonequilibrium. Notably, in the case of the equilibrium motion of a Brownian particle in a confining potential, we show that $P(t_{\rm m}|T)$ becomes universal at late times. Moreover, we demonstrate that for any equilibrium process the PDF $P(t_{\rm m}|T)$ is symmetric around the midpoint $t_{\rm m}=T/2$. For two nonequilibrium processes, namely the resetting BM and a single confined RTP, we verify by computing $P(t_{\rm m}|T)$ exactly that this symmetry is not present (see Fig. \ref{fig:small_t}). Thus, the measurement of the distribution of $t_{\rm m}$ provides a simple recipe to detect nonequilibrium dynamics in a stationary time series.


We start by investigating $P(t_{\rm m}|T)$ in the case of equilibrium systems. The process that we consider is an overdamped $1d$ Brownian particle (with diffusion coefficient $D$ and friction coefficient $\Gamma=1$) moving in a symmetric confining potential that grows for large $|x|$ as $V(x) \approx \alpha |x|^p$, where $\alpha>0$ and $p>0$. In this case, the system has an equilibrium stationary state characterized by the Gibbs-Boltzmann measure $P_{\rm st}(x)\simeq e^{-V(x)/D}$ where $D$ is exactly the temperature by fluctuation dissipation theorem. Computing the distribution of $t_{\rm m}$ for any $p$ is challenging. However, in the special cases $p=1$ and $p=2$, we are able to compute exactly $P(t_{\rm m}|T)$ \cite{supmat}. For instance, for $V(x)=\alpha x^2$ (corresponding to the Ornstein-Uhlenbeck process), we obtain $P(t_{\rm m}|T)=\alpha F_{\rm OU}[\alpha t_{\rm m},\alpha (T-t_{\rm m})]$, where the double Laplace transform of the scaling function $F_{\rm OU}(t_1,t_2)$ is given by
\begin{eqnarray}\label{integral_OU}
&&\int_{0}^{\infty}dt_1\int_{0}^{\infty}dt_2~F_{\rm OU}(t_1,t_2)e^{-s_1t_1-s_2t_2}\\ &=&
\frac{1}{2\sqrt{2\pi}}\int_{-\infty}^{\infty}dz~e^{-z^2/2}\frac{D_{-1-s_1/2}(z)}{D_{-s_1/2}(z)}\frac{D_{-1-s_2/2}(z)}{D_{-s_2/2}(z)}\,, \nonumber
\end{eqnarray}
where $D_{\nu}(z)$ is the parabolic-cylinder function \cite{grads}. Using Eq. \eqref{integral_OU}, it is easy to see that $F_{\rm OU}(t_1,t_2)=F_{\rm OU}(t_2,t_1)$, leading to $P(t_{\rm m}|T)=P(T-t_{\rm m}|T)$. As we demonstrate below, the symmetry of $P(t_{\rm m}|T)$ around the midpoint $t_{\rm m}=T/2$ is a general feature of equilibrium processes. This property is a direct consequence of the time reversibility, which is always present at equilibrium. Conversely, if the process is out-of-equilibrium, the distribution of $t_{\rm m}$ might or might not be symmetric.

The PDF $P(t_{\rm m}| T)$ can be also derived exactly for $p=1$ (see Eq. (3) of~\cite{supmat}). In addition to proving the symmetry $P(t_{\rm m}| T) =  P(T-t_{\rm m}| T)$, the exact result in Eq. (\ref{integral_OU}) for $p=2$, and the analogous one for $p=1$ (see Eq. (8) in \cite{supmat}), can be used to extract the asymptotic behaviour of $P(t_{\rm m}| T)$ in the large $T$ limit. In this limit, we find that there is a ``bulk'' regime where $P(t_{\rm m}| T)\approx 1/T$ is essentially flat. However, near the two ``edges'' $t_{\rm m} = 0$ and $t_{\rm m}=T$ (symmetrically), the distribution $P(t_{\rm m}| T)$ has a nontrivial shape [see Fig. \ref{fig:small_t}~a)]. Moreover, near the edges, once appropriately scaled, the scaling form of $P(t_{\rm m}| T)$ turns out to be identical for both $p=1$ and $p=2$! This ``universality'' is rather unexpected and naturally leads us to wonder
whether the edge behavior of $P(t_{\rm m}| T)$ for general $p>0$ is also universal. We show that indeed this universality holds for any $p \geq 1$, i.e., for sufficiently confining potentials. However, for $0<p<1$, i.e. for ``shallow'' potentials, there is no universal edge behavior.

\begin{figure}[t]
\includegraphics[scale=0.6]{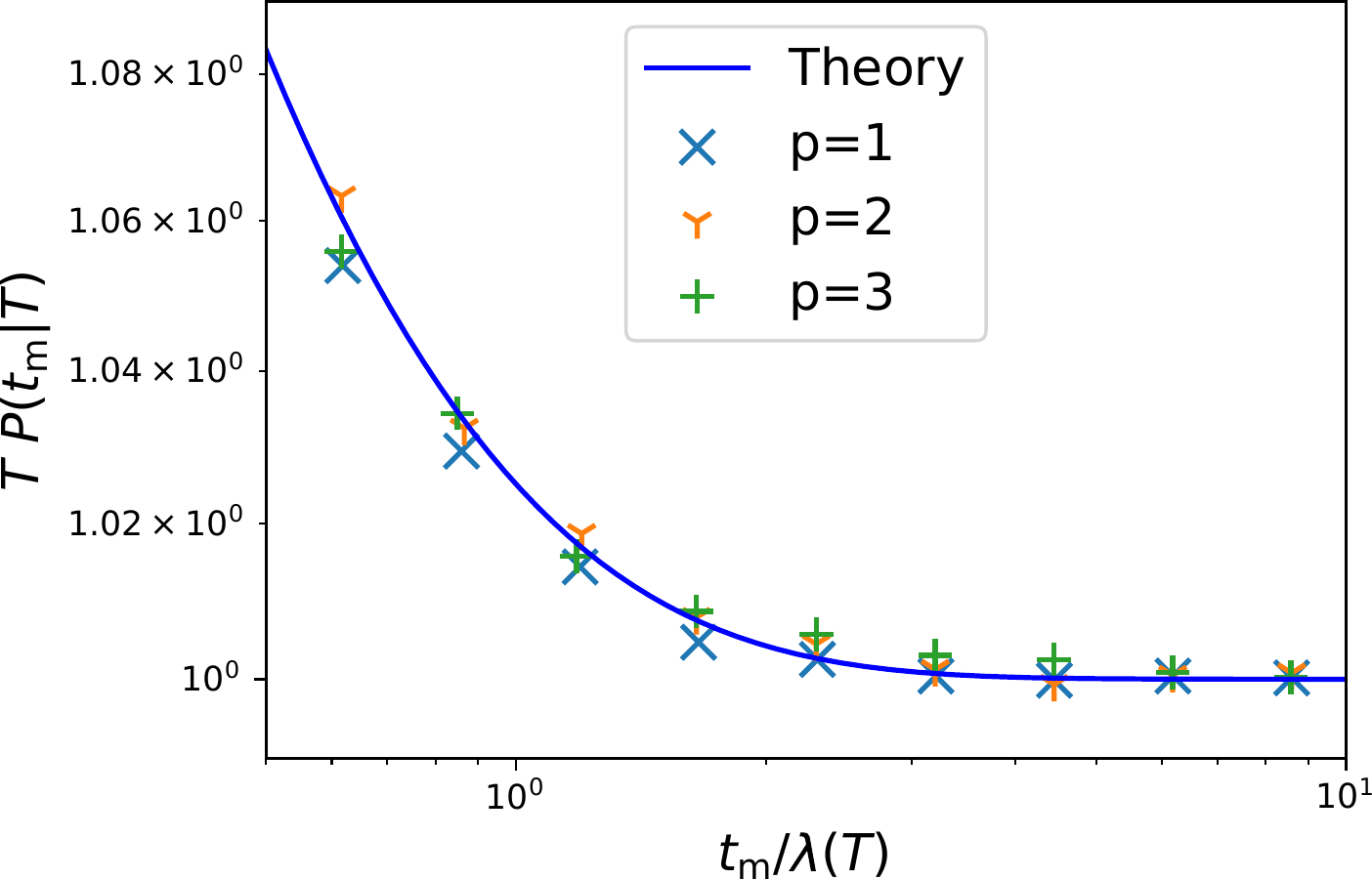} 
\caption{\label{fig:comparison} The scaled distribution $T P(t_{\rm m}|T)$ as a function of $t_{\rm m}/\lambda(T)$ in the region $t_{\rm m}\in [0,10~\lambda(T)]$. The symbols depict the results of numerical simulations with the potential $V(x)=|x|^p$, with $p=1,2$ and $3$, and large $T$ ($T=6400$ for $p=1$ and $T=800$ for $p=2,3$). The continuous line corresponds to the analytical result in Eqs. \eqref{universal_limit} and (\ref{G}). }
\end{figure}

In the absence of an exact result for general $p \geq 1$, we develop a real-space ``blocking argument'' (\`a la Kadanoff), which demonstrates clearly this universality of the edge behavior for $p\geq 1$. More precisely, we find that, for $p\geq 1$,
\begin{equation}
P(t_{\rm m}|T)\simeq \begin{cases}
\frac{1}{T} G\left[\frac{t_{\rm m}}{\lambda(T)}\right]~&\text{for}~ t_{\rm m}\lesssim \lambda(T),\\
\frac{1}{T}~&\text{for}~\lambda(T)\ll t_{\rm m} \ll T-\lambda(T), \\
\frac{1}{T} G\left[\frac{T-t_{\rm m}}{\lambda(T)}\right]~&\text{for}~t_{\rm m}\gtrsim T-\lambda(T)\,, \\
\end{cases}
\label{universal_limit}
\end{equation}
with $\lambda(T)=\frac{4D}{\alpha^2 p^2}\left[\frac{D}{\alpha}\log(T)\right]^{-2(p-1)/p}$ denoting the width of the 
edge region and the universal scaling function is given by
\begin{equation}
G(z)=\frac{1}{2}\left[1+\frac{e^{-z}}{\sqrt{\pi z}}+\operatorname{erf}(\sqrt{z})\right]\,,
\label{G}
\end{equation}
where ${\rm erf}(x) = (2/\sqrt{\pi}) \, \int_0^x e^{-u^2} \, du$. The late time distribution $P(t_{\rm m}|T)$ in Eq. (\ref{universal_limit}) is manifestly symmetric around $t_{\rm m} = T/2$ for all $p \geq 1$ and the dependence on the parameters $p$ and $\alpha$ appears only through the width $\lambda(T)$. 
When $z\to 0$, $G(z)$ diverges as $1/(2 \sqrt{\pi z})$. On the other hand, for large $z$, $G(z)$ goes to the limit value $1$, smoothly connecting with the central part where $P(t_{\rm m}|T)\simeq 1/T$. The scaled distribution $TP(t_{\rm m}|T)$ is shown as a function of $t_{\rm m}/\lambda(T)$ in Fig. \ref{fig:comparison}. The numerical curves obtained for different values of $p$ collapse onto the same theoretical curve, given in Eq. \eqref{universal_limit}. We have checked that the deviations from the theoretical curve are a consequence of finite-size effects \cite{supmat}. Note that for $p=1$ the width $\lambda(T)$ is a constant independent of $T$, while for $p>1$ it shrinks as $\log(T)^{-2(p-1)/p}$ for large $p$.

We next focus on nonequilibrium stationary processes. One of the simplest nonequilibrium models is BM with stochastic resetting \cite{EM11,EMS20}. Here, we consider a one-dimensional BM, whose position is reset to the origin randomly in time with constant rate $r$. The resetting dynamics induces a nonzero net probability current towards the origin, driving the system to a nonequilibrium
stationary state where the position distribution, in $1d$, is known to be $P_{\rm st}(x)=\sqrt{r/(4D)}e^{-\sqrt{r/D}|x|}$, where $D$ is the diffusion constant \cite{EM11}. 
The distribution $P(t_{\rm m}|T)$ for this process has been recently studied where the starting position is fixed~\cite{SP21}. Here, instead, we assume that the initial position of the particle is drawn from the stationary state $P_{\rm st}(x)$. We show that $P(t_{\rm m}|T)=r F_{\rm R}[rt_{\rm m},r(T-t_{\rm m})]$, where the scaling function $F_{\rm R}(t_1,t_2)$ is given in Eq. (8) of~\cite{supmat}. In this case, we find that $F_{\rm R}(t_1,t_2)\neq F_{\rm R}(t_2,t_1)$, implying that $P(t_{\rm m}|T)$ is not symmetric around $t_{\rm m}=T/2$. This asymmetry is confirmed by numerical simulations (see Fig. \ref{fig:small_t}b) and analytically (see Eqs. (9) and (10) as well as Fig. 2 in \cite{supmat}).

The second nonequilibrium process that we consider is a single RTP with fixed velocity $v_0$, moving in a one-dimensional potential $V(x)=\alpha |x|$, with $\alpha>v_0$ (for the details of the model, see \cite{supmat}). In the context of active matter, the RTP model has been widely studied \cite{C12,HJ1995,berg_book,TC2008,FM2018,DKM19}. We compute exactly $P(t_{\rm m}|T)$ for this model, showing that it is not symmetric around $t_{\rm m}=T/2$ \cite{supmat}.

\begin{figure}[t]
\includegraphics[scale=0.65]{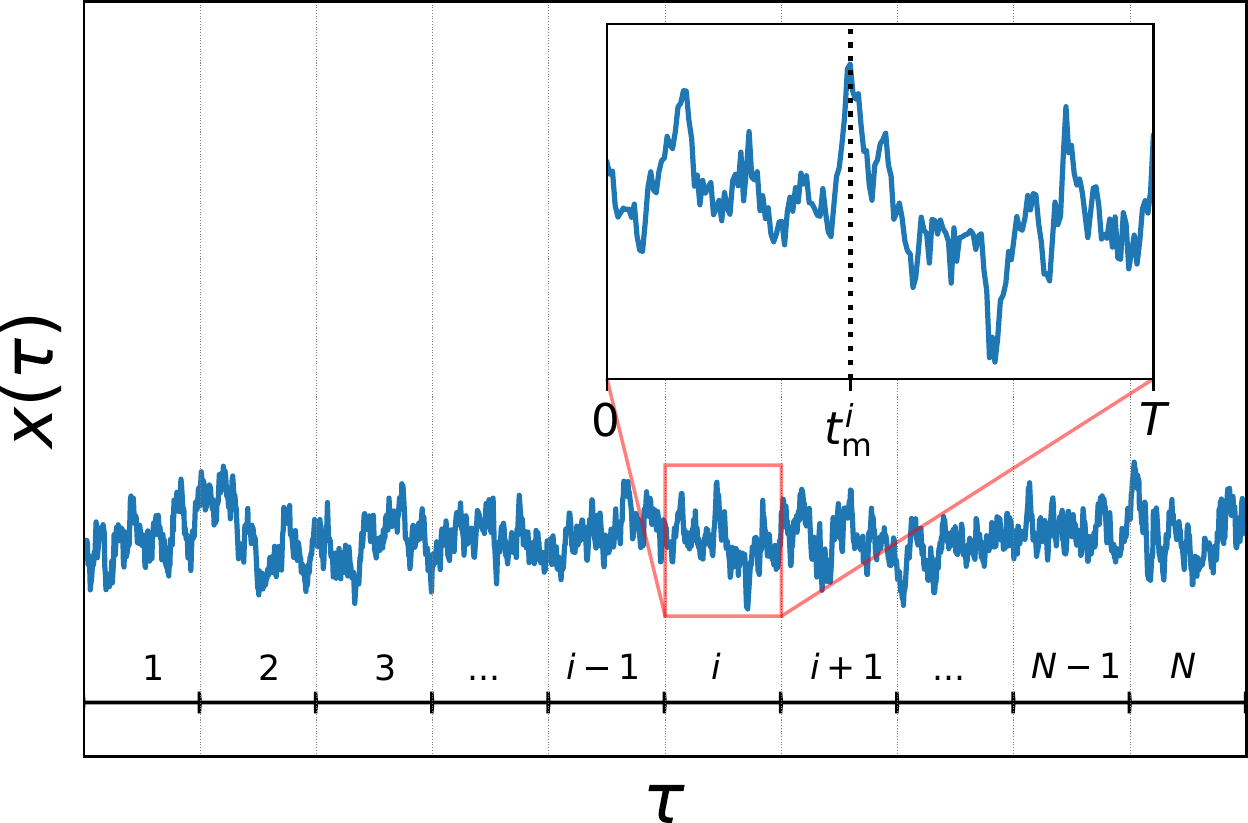} 
\caption{\label{fig:trajectory_old} Long stationary process $x(\tau)$ as a function of time $\tau$. The trajectory can be divided into $N$ blocks of duration $T$. For $1\leq i\leq N$, we compute the time $t_{\rm m}^i$ at which the maximum of $x(\tau)$ within the $i$-th block is reached (see inset). From the histogram of these $N$ variables $t_{\rm m}^1\,,\ldots\,,t_{\rm m}^N$, one can estimate the distribution $P(t_{\rm m} \vert T)$, defined in the text. If $P(t_{\rm m} \vert T)$ is not symmetric around $t_{\rm m}=T/2$, then the process is out-of-equilibrium.}
\end{figure}

Interestingly, the fact that for all equilibrium processes the distribution $P(t_{\rm m}|T)$ is symmetric around $t_{\rm m}=T/2$ provides a simple criterion to detect nonequilibrium dynamics in stationary time series. More precisely, imagine that one has access only to a long stationary time series $x(\tau)$ as a function of time $\tau$, e.g., from experimental measurements (see Fig. \ref{fig:trajectory_old}), but with no other additional information. This setup is motivated by the increasing interest in single-particle tracking, which provides individual-particle trajectories with high space-time resolution \cite{BJS04,BDM12,MJC14,KDA18,BBP20,TPS20,BFP21}. For instance, this time series $x(\tau)$ could represent the location of a confined active particle or the position of a BM in an optical trap. Then, a natural question arises: is there a simple way to determine whether or not $x(\tau)$ is at equilibrium, without any a priori knowledge of its underlying dynamics? In recent years, several attempts to answer this question have been made \cite{GMG18}. One possibility is the verification of the so-called fluctuation-dissipation theorem, which is only valid at equilibrium \cite{CDK1997,MHJ01,MTS07,TFA16,GMG18}. As an example, this method has been employed to show the nonequilibrium nature of red blood cells \cite{TFA16}. Several other methods, based, e.g., on the detection of probability currents in the phase space or the breakdown of time-reversal symmetry, have also been developed \cite{ZS07,RP16,GFM16,GMG18,WFM20,Byrne21,FJC21,LH19,MBH19,MGK20,HG20,OID20,OMS20,SMM21,WFM20,SZ14,MMZ16,ZWM16,MMZ17,BBF16,GFM16,KCB18}. 

Here, we propose the following simple recipe which consists of two steps. a) Divide the long time series $x(\tau)$ into $N$ blocks each of duration $T$ (see Fig. \ref{fig:trajectory_old}) and measure the time $t_{\rm m}^i$ at which the maximum occurs within the $i$-th block. From the histogram of the $N$ values ${t}^1_{\rm m}\,,\ldots,{t}^N_{\rm m}$, one then constructs the empirical PDF $P(t_{\rm m}|T)$, where $0\leq t_{\rm m}\leq T$. b) Check if the empirical PDF $P(t_{\rm m}|T)$ is symmetric around $t_{\rm m}=T/2$. Our test predicts that if $P(t_{\rm m}|T)$ is asymmetric around $T/2$ [as in Fig. \ref{fig:small_t}b)], the dynamics of $x(\tau)$ is nonequilibrium. Conversely, if $P(t_{\rm m}|T)$ is symmetric around $t_{\rm m}=T/2$ [as in Fig. \ref{fig:small_t}a)], our test is inconclusive and one has to resort to more sophisticated techniques. The asymmetry in $P(t_{\rm m}|T)$ is a clear signature of the nonequilibrium nature of $x(\tau)$. Our test is also applicable to systems composed of several interdependent variables. For such cases, using our criterion, finding that the distribution of $t_{\rm m}$ for any one of these variables is not symmetric around $T/2$ is sufficient to determine that the full system is out of equilibrium.

Let us also mention that there exist nonequilibrium processes for which our criterion is inconclusive. For instance, let us consider another model of active matter, namely a single one-dimensional active Ornstein-Uhlenbeck particle (AOUP) in a harmonic potential \cite{FNC16,DBE20}. It is possible to show that in this case, despite the system being nonequilibrium, the distribution $P(t_{\rm m}|T)$ of the time of the maximum is symmetric around $t_{\rm m}=T/2$ \cite{DBE20}. This is just a consequence of the fact that the AOUP in a harmonic potential is a Gaussian stationary process. Indeed, it is possible to show that for any Gaussian stationary process the distribution of $t_{\rm m}$ is symmetric around~$T/2$~\cite{supmat}.

We start by sketching the blocking argument that leads to the universal result in Eqs. (\ref{universal_limit}) and (\ref{G}) for all $p \geq 1$. We consider the position $x(\tau)$ of a single overdamped Brownian particle in a confining potential growing as $V(x)\simeq \alpha |x|^p$ for large $|x|$, with $\alpha>0$ and $p\geq 1$. The Langevin equation that describes the evolution of $x(\tau)$ is
\begin{equation}
\frac{dx(\tau)}{d\tau}=-V'(x)+\eta(\tau)\,,
\label{langevin}
\end{equation}
where $\eta(\tau)$ is Gaussian white noise with zero mean and correlator $\langle \eta(\tau)\eta(\tau')\rangle=2D\delta(\tau-\tau')$ and $V'(x)=dV(x)/dx$. For $p\geq 1$, one can show that the autocorrelation function $\langle x(\tau)x(\tau ')\rangle -\langle x(\tau)\rangle\langle x(\tau')\rangle $ decays exponentially in $|\tau-\tau'|$ over a typical time $T_B\sim O(1)$ \cite{SM2020}. For $T\gg T_B$, we can divide the time interval $[0,T]$ into $N_B$ blocks of identical size $T_B$, which are essentially uncorrelated. Let $m_i$ be the maximal position reached in the $i$-th block.
Clearly the variables $m_i$'s are independent of each other (since they belong to different blocks), but they are identically distributed due to the stationarity of the process. 
This implies that the probability that the maximum is reached in the $i$-th box is the same for each box and thus it is simply $1/N_B=T_B/T$. This argument suggests that the probability distribution of $t_{\rm m}$ is approximately given by the uniform measure $P(t_{\rm m}|T)\simeq 1/T$. However, this argument is only valid in the bulk of the distribution $P(t_{\rm m}|T)$, i.e., when $T_B\ll t_{\rm m}\ll T-T_{B}$. In the regions $0<t_{\rm m}<T_B$ and $T-T_B<t_{\rm m}<T$, a detailed analysis, taking into account edge effects, is required. 

To show this, we consider the interval $[0,T_B]$ and condition on the event that the maximum is reached in this first block. Since the position $x(\tau)$ in this block will be very close to the maximal position $M$, we can linearize the potential $V(x)$ around $x=M$. To leading order, the Langevin equation \eqref{langevin} becomes
\begin{equation}
\frac{dx(\tau)}{d\tau}=-V'(M)+\eta(\tau)\,.
\label{langevin2}
\end{equation}
In first approximation, the particle is subject to a constant negative drift $\mu=-V'(M)<0$. For large $T$, the maximum $M$ typically grows as $\left(\frac{D}{\alpha}\log(T)\right)^{1/p}$ \cite{supmat}. Consequently, the constant drift $\mu$ is given by
\begin{equation}
\mu\simeq-\alpha~ p\left(\frac{D}{\alpha}\log(T)\right)^{(p-1)/p}\,.
\label{mu}
\end{equation}

The PDF of the time $t_{\rm m}$ of the maximum in a time interval $[0,T_B]$ of a BM with constant drift $\mu$ has been computed in Ref. \cite{MB08} and is given by
\begin{equation}
P(t_{\rm m}|T_B)=\frac{h_{\mu}(t_{\rm m})h_{-\mu}(T_B-t_{\rm m})}{\pi\sqrt{t_{\rm m}(T_B-t_{\rm m})}}\,,
\label{PTb}
\end{equation}
where 
\begin{equation}
h_{\mu}(\tau)=e^{-\mu^2 \tau/4D}+ \mu \sqrt{\frac{\pi  \tau}{4D}}\left[1+ \operatorname{erf}\left( \mu \sqrt{\frac{\tau}{4D} }\right)\right]\,.
\label{eq:h_mu}
\end{equation}
Thus, for $0\leq t_{\rm m}\ll 1$ and $T\gg 1$, the distribution of $t_{\rm m}$ can be written as
\begin{equation}
P(t_{\rm m}|T)\simeq \frac{T_B}{T}\frac{h_{\mu}(t_{\rm m})h_{-\mu}(T_B-t_{\rm m})}{\pi\sqrt{t_{\rm m}(T_B-t_{\rm m})}}\,,
\label{eq:edge1_drift}
\end{equation}
where the drift $\mu$ is given in Eq. \eqref{mu}. We recall that the term $T_B/T$ is the probability that the maximum falls in the first block. Note that, since we do not know the precise value of $T_B$, the result in Eq. \eqref{eq:edge1_drift} gives us the edge behavior of $P(t_{\rm m}|T)$ up to a multiplicative constant. In particular, in the region where $t_{\rm m}\ll T_B$, we obtain
\begin{equation}
P(t_{\rm m}|T)\propto \frac{1}{T}\frac{h_{\mu}(t_{\rm m})}{\sqrt{t_{\rm m}}}\,.
\label{eq:edge1_drift2}
\end{equation}
Finally, the multiplicative factor can be obtained by imposing that the edge expression in Eq. \eqref{eq:edge1_drift2} matches for large $t_{\rm m}$ with the bulk result $P(t_{\rm m}|T)\simeq 1/T$ and, using the expression of $h_{\mu}(\tau)$ in Eq. \eqref{eq:h_mu}, we obtain the result in Eq. \eqref{universal_limit}. An analogous derivation can be carried out for the right edge of $P(t_{\rm m}|T)$. In the special cases $p=1$ and $p=2$, where we could compute $P(t_{\rm m}|T)$ exactly~\cite{prep}, the asymptotic analysis for large $T$ is fully consistent with the approximate block argument developed above for arbitrary $p \geq 1$. Note that in the case $0<p<1$ the result in Eq. \eqref{universal_limit} is not valid since the autocorrelation function of $x(\tau)$ does not decay exponentially in time \cite{SM2020}.

We next present the derivation of the fact that $P(t_{\rm m}|T)$, for any equilibrium stationary process on the interval $[0,T]$, is symmetric around $t_{\rm m} = T/2$. For simplicity, we consider a discrete-time process $x_k$, with $1\leq k\leq T$. It is easy to generalize the following derivation to continuous time. Note that here $t_{\rm m}$ and $T$ are integer numbers. Denoting by $P\left(\{x_k\}\right)$ the probability of observing the trajectory $\{x_k\}=\{x_1,\ldots ,x_T\}$, the distribution of the time $t_{\rm m}$ of the maximum can be written as
\begin{equation}
P(t_{\rm m}|T)=\int_{-\infty}^{\infty}dx_1\ldots\int_{-\infty}^{\infty}dx_T~  \Theta_{t_m}(\{x_k\}) P\left(\{x_k\}\right)\,,
\label{discrete}
\end{equation}
where $\Theta_{k}\left(\{x_i\}\right)=\prod_{i\neq k}\theta\left(x_k-x_i\right)$ and $\theta(z)$ is the Heaviside step function, i.e., $\theta(z)=1$ for $z>0$ and $\theta(z)=0$ otherwise. In other words, $\Theta_{k}\left(\{x_i\}\right)$ is one if the maximum of the trajectory $\{x_i\}$ is reached at step $k$ and zero otherwise. Thus, in Eq. \eqref{discrete}, we integrate over all possible trajectories for which the time of the maximum is $t_{\rm m}$. Let us denote by $\{\bar{x}_k\}=\{x_{T-k}\}$ the time-reversed trajectory associated to $\{x_{k}\}$. For an equilibrium process, it is possible to show that, as a consequence of the detailed balance condition, $P(\{x_k\})=P(\{\bar{x}_k\})$ (this is not true in general for nonequilibrium processes). Using this result in Eq. \eqref{discrete} and performing the change of variables $x_i\to\bar{x}_i=x_{T-i}$, we obtain 
\begin{equation}
P(t_{\rm m}|T)=\int_{-\infty}^{\infty}d\bar{x}_1\ldots\int_{-\infty}^{\infty}d\bar{x}_T  ~\Theta_{t_m}(\{\bar{x}_{T-k}\}) P\left(\{\bar{x}_k\}\right)\,.
\label{discrete2}
\end{equation}
It is easy to show that $\Theta_{t_m}(\{\bar{x}_{T-k}\})=\Theta_{T-t_m}(\{\bar{x}_{k}\})$ and thus we find
\begin{equation}
P(t_{\rm m}|T)=\int_{-\infty}^{\infty}d\bar{x}_1\ldots\int_{-\infty}^{\infty}d\bar{x}_T  ~\Theta_{T-t_m}(\{\bar{x}_{k}\}) P\left(\{\bar{x}_k\}\right)\,.
\label{discrete3}
\end{equation}
Recalling the expression for $P(t_m|T)$, given in Eq. \eqref{discrete}, we obtain our desired result $P(t_m|T)=P(T-t_m|T)$, which is thus a necessary, but not a sufficient, condition for a stationary process to be at equilibrium.


To conclude, we have investigated the distribution $P(t_{\rm m}|T)$ of the time $t_{\rm m}$ at which a stationary process of duration $T$ reaches its global maximum. Using path integral techniques, we have computed exactly $P(t_{\rm m}|T)$ for several stationary processes. In particular, for a diffusive particle in a trapping potential, we have further shown that $P(t_{\rm m}|T)$, suitably scaled, is universal at late times, i.e., independent of the details of the potential. Moreover, we have presented a simple sufficiency test to detect whether a stationary time series has nonequilibrium dynamics. Our method is based on estimating the PDF $P(t_{\rm m}|T)$. If it is asymmetric the dynamics is necessarily nonequilibrium. The test proposed in this Letter is very general and can be applied to any stationary process.

\acknowledgments We thank R. K. P. Zia for useful discussions.

\newpage

.

\newpage

\setcounter{secnumdepth}{2}

\begin{widetext}
\begin{Large}
\begin{center}

Supplementary Material for\\  {\it Distribution of the time of the maximum for stationary processes}

\end{center}
\end{Large}

\bigskip

\section{Distribution of the time of the maximum: summary of the main results}

In this Section, we present the main results on the distribution of the time $t_{\rm m}$ of the maximum for the stationary processes considered in the Letter. The corresponding formulae are too long to be included in the main text and are thus presented here. The details of the computations, performed using path-integral techniques, are quite long and will be presented elsewhere \cite{prep}. Here, with the help of Fig. \ref{fig:trajectory_decomposition}, we sketch this path-integral method, which can be summarized in the following steps:
\begin{enumerate}
\item divide the time interval $[0,T]$ into two subintervals $[0,t_{\rm m}]$ (I) and $[t_{\rm m},T]$ (II),
\item in the first interval (I) the process starts from position $x_0$, drawn from the stationary measure $P_{\rm st}(x_0)$, and reaches the maximum $M$ without crossing position $M$, 
\item in the second interval (II) the process starts from position $M$ and reaches the final position $x_F<M$ without crossing position $M$.
\end{enumerate}
Thus, the joint probability of $x_0$, $M$, $t_{\rm m}$, and $x_F$ can be written as
\begin{equation}
P(x_0,M,x_F,t_{\rm m}|T)=P_{\rm st}(x_0)G_M(M,t_{\rm m}|x_0,0)
G_M(x_F,T|M,t_{\rm m})\,,\label{joint}
\end{equation}
with the constraint $x_0<M$ and $x_F<M$. In Eq. (\ref{joint})
$G_M(x_2,t_2|x_1,t_1)$ is the {\it constrained} propagator of the process, 
defined as the probability that the process arrives 
at position $x_2$ at time 
$t_2$ starting from position $x_1$ at time $t_1$ with $t_2\ge t_1$, while staying below $M$ in the time interval 
$(t_1,t_2)$. Note that to obtain Eq. \eqref{joint} we have assumed that the process is Markov. This technique can be also generalized to some non-Markov processes, e.g. run-and-tumble particles (RTPs). Integrating over $x_0$, $M$, and $x_F$, we find that the distribution of $t_{\rm m}$ can be written as
\begin{equation}
P(t_{\rm m}|T)=\int_{-\infty}^{\infty}dx_0~\int_{x_0}^{\infty}
dM~\int_{-\infty}^{M}dx_F~P_{\rm st}(x_0)\,
G_M(M,t_{\rm m}|x_0,0)\,G_M(x_F,T|M,t_{\rm m})\,.
\label{joint_int.1}
\end{equation}
Thus, in order to compute $P(t_{\rm m}|T)$ one needs the constrained 
propagator $G_M(x_2,t_2|x_1,t_1)$ , which is usually not easy to 
compute. However, for some processes, one can compute the constrained
propagator exactly as mentioned below. One can then substitute this 
constrained propagator in Eq. (\ref{joint_int.1})-- however
performing this triple integral explicitly is highly nontrivial.
In most solvable cases shown below, it however allows us
to obtain an exact expression for the double Laplace transform
of $P(t_m|T)$ (see, for example, Eq. (\ref{integral_OU}) in the main text).

\subsection{Brownian motion in a confining potential}

The equilibrium process we consider is an overdamped Brownian particle in a confining potential $V(x)=\alpha |x|^p$, with $\alpha>0$ and $p>0$. The position $x(\tau)$ of the particle evolves according to the Langevin equation
\begin{equation}
\frac{dx(\tau)}{d\tau}=-V'(x)+\eta(\tau)\,,
\label{sup:langevin}
\end{equation}
where $\eta(\tau)$ is a Gaussian white noise with zero mean and correlator $\langle \eta(\tau)\eta(\tau')\rangle=2D\delta(\tau-\tau')$ and $V'(x)=dV(x)/dx$. The equilibrium stationary state of this system is given by $P_{\rm st}(x)\propto e^{-V(x)/D}$. We assume that the particle starts at the initial time from some position $x_0$, drawn from the equilibrium distribution $P_{\rm st}(x_0)$, and that it evolves according to Eq. \eqref{sup:langevin} up to time $T$. We are interested in computing the distribution $P(t_{\rm m }|T)$ of the time $t_{\rm m}$ at which the position of the particle reaches its maximal value up to time $T$. Computing $P(t_{\rm m }|T)$ for any $p>0$ is challenging. However, we compute exactly $P(t_{\rm m }|T)$ in the special cases $p=1$ and $p=2$. 

In the case $p=1$, we show that \cite{prep}
\begin{equation}\label{F1}
P(t_{\rm m}|T)=\frac{\alpha^2}{4D}F_1\left[\frac{\alpha^2}{4D}t_{\rm m},\frac{\alpha^2}{4D}(T-t_{\rm m})\right]\,,
\end{equation}
where the double Laplace transform of the scaling function $F_1(t_1,t_2)$ is given by
\begin{eqnarray}\label{eq:LT_F_V}
&&\int_{0}^{\infty}dt_1 \int_{0}^{\infty}dt_2~F_1(t_1,t_2)e^{-s_1t_1-s_2t_2}\\&=& \frac{1}{2(1+\sqrt{1+s_1})(1+\sqrt{1+s_2})}\Bigg[1+\int_{0}^{\infty}dz\,e^{-z}\frac{\left(\sqrt{1+s_1}+1-e^{-\sqrt{1+s_1}z}\right)\left(\sqrt{1+s_2}+1-e^{-\sqrt{1+s_2}z}\right)}{\left(\sqrt{1+s_1}-1+e^{-\sqrt{1+s_1}z}\right)\left(\sqrt{1+s_2}-1+e^{-\sqrt{1+s_2}z}\right)}\Bigg]\,.\nonumber
\end{eqnarray}
From Eq. \eqref{eq:LT_F_V}, it is easy to check that $P(t_{\rm m}|T)$ is correctly normalized to unity. Inverting the double Laplace transform in Eq. \eqref{eq:LT_F_V} is highly nontrivial. However, it is easy to check that $F_1(t_1,t_2)=F_1(t_2,t_1)$ and hence $P(t_{\rm m}|T)=P(T-t_{\rm m}|T)$, in agreement with the fact that the process is at equilibrium. Consequently, the first moment of $t_{\rm m}$ is simply given by $\langle t_{\rm m}\rangle=T/2$.

\begin{figure}[t]
\includegraphics[scale=0.7]{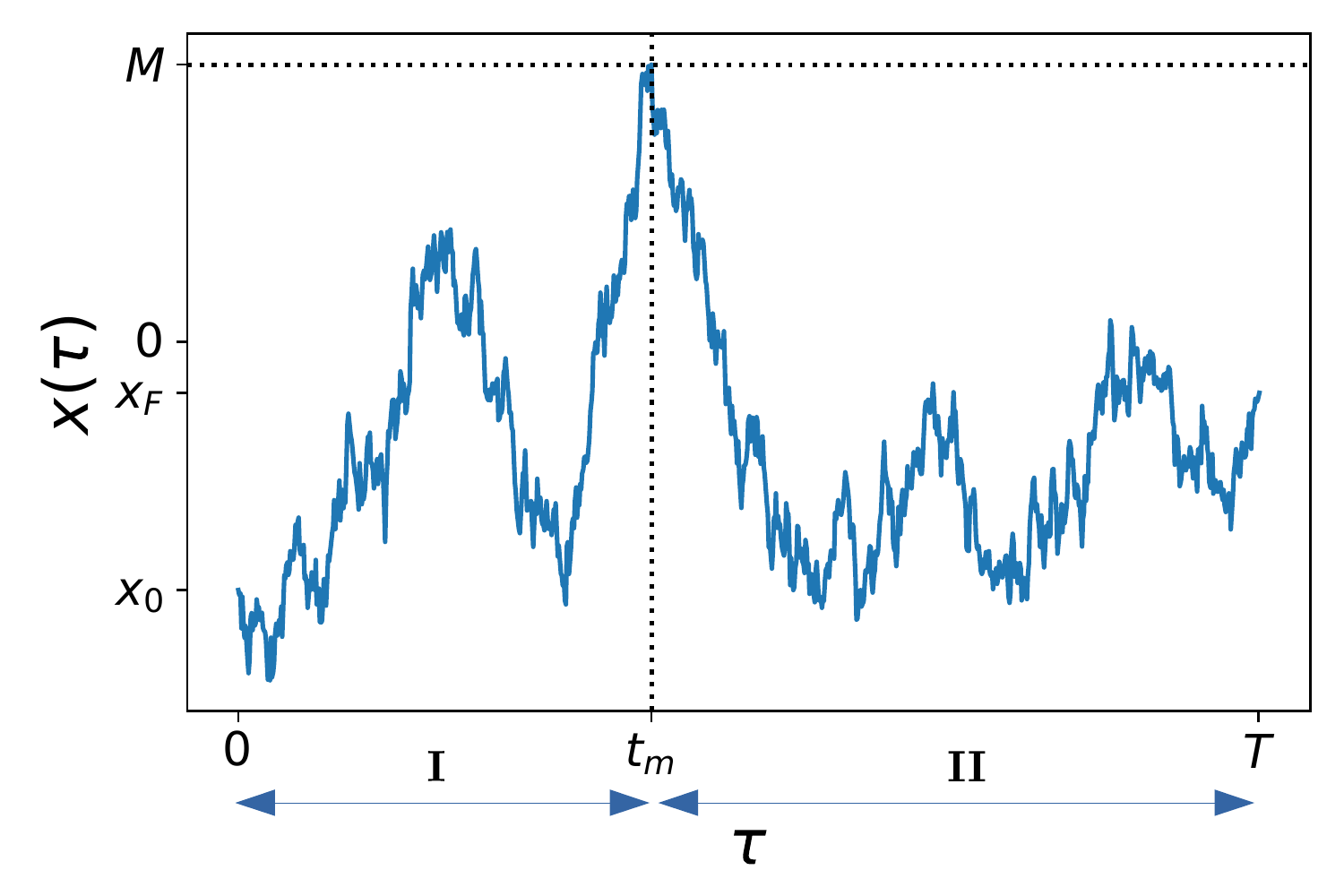} 
\caption{\label{fig:trajectory_decomposition} Decomposition of a stationary trajectory $x(\tau)$ of duration $T$. In the first interval $[0,t_{\rm m}]$ (I) the process starts from position $x_0$, it remains below position $M$, and reaches the maximum $M$ at time $t_{\rm m}$. In the second interval $[t_{\rm m},T]$, it starts from position $M$ at time $t_{\rm m}$ and it reaches position $x_F$ at time $T$, without crossing position $M$.}
\end{figure}

In the case $p=2$, corresponding to the Ornstein-Uhlenbeck process, we find that 
\begin{equation}\label{F2}
P(t_{\rm m}|T)=\alpha F_{\rm OU}\left[\alpha t_{\rm m},\alpha (T-t_{\rm m})\right]
\end{equation}
where the double Laplace transform of the scaling function $F_{\rm OU}(t_1,t_2)$ is given by
\begin{equation}
\int_{0}^{\infty}dt_1\int_{0}^{\infty}dt_2~F_{\rm OU}(t_1,t_2)e^{-s_1t_1-s_2t_2}=
\frac{1}{2\sqrt{2\pi}}\int_{-\infty}^{\infty}dz~e^{-z^2/2}\frac{D_{-1-s_1/2}(z)}{D_{-s_1/2}(z)}\frac{D_{-1-s_2/2}(z)}{D_{-s_2/2}(z)}\,,\
\label{supmat:integral_OU}
\end{equation}
as given in Eq. (1) in the main text. Here $D_{\nu}(z)$ is the parabolic-cylinder function \cite{grads}. From Eq. \eqref{supmat:integral_OU}, it is possible to check that $P(t_{\rm m}|T)$ is normalized to unity. Moreover, it is easy to show that $F_{\rm OU}(t_1,t_2)=F_{\rm OU}(t_2,t_1)$ and consequently $P(t_{\rm m}|T)=P(T-t_{\rm m}|T)$. Hence, the first moment of $t_{\rm m}$ is simply given by $\langle t_{\rm m}\rangle=T/2$.

Taking the limit $T\to \infty$ in Eqs. \eqref{F1} and \eqref{F2}, it is possible to show that the exact expressions for $P(t_{\rm m}|T)$, valid for $p=1$ and $p=2$, converge to the universal form given in Eq. (2) of the main text, valid for $p\geq 1$. This asymptotic result is also verified numerically for $p=1,2,$ and $p=3$, as shown in Fig. \ref{fig:convergence}. We observe that the numerical curves approach the analytic result as $T$ increases.

\begin{figure*}[t]
\includegraphics[scale=0.4]{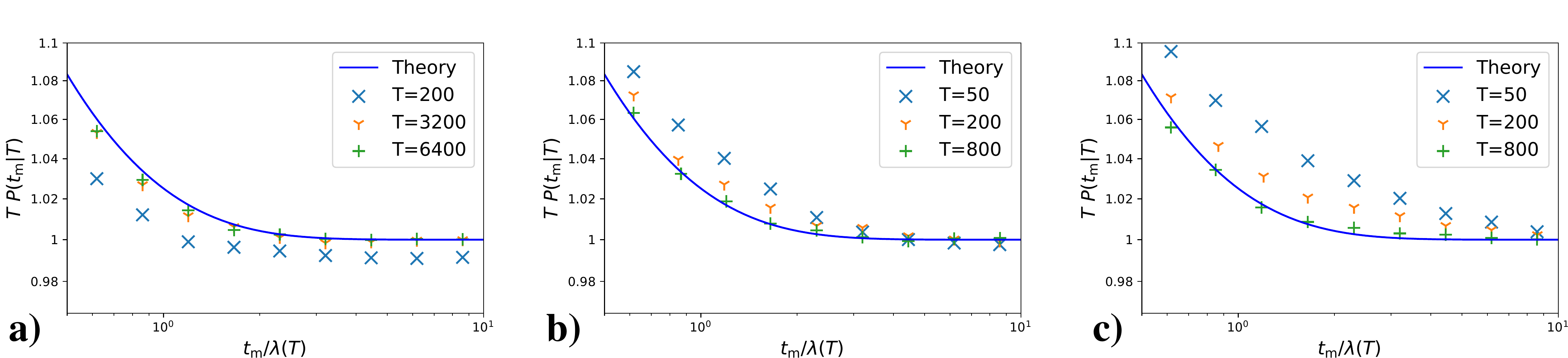} 
\caption{\label{fig:convergence} The scaled distribution $T P(t_{\rm m}|T)$ as a function of $t_{\rm m}/\lambda(T)$ for different values of $T$ and for $p=1$ (panel {\bf a}), $p=2$ (panel {\bf b}), and $p=3$ (panel {\bf c}). The symbols depict the results of numerical simulations while the continuous lines correspond to the analytical result in Eq. (2) of the main text, valid in the limit of large $T$.}
\end{figure*}

\subsection{Brownian motion with resetting}

The first out-of-equilibrium process that we consider is Brownian motion (BM) with stochastic resetting, which has been extensively studied in recent years \cite{EM11,EMS20}. We consider a Brownian particle, diffusing in one dimension with diffusion constant $D$ and resetting to the origin with constant rate $r$. It is possible to show that the system admits the following nonequilibrium steady state \cite{EM11}
\begin{equation}
P_{\rm st}(x)=\sqrt{\frac{r}{4D}}\exp\left(-\sqrt{\frac{r}{D}}|x|\right)
\label{stationary_res}
\end{equation}
We assume that the particle starts from position $x_0$, drawn from the stationary state \eqref{stationary_res}, and that it evolves up to time $T$. We show that the probability density function (PDF) $P(t_{\rm m}|T)$ of the time $t_{\rm m}$ at which the particle reaches its maximal position up to time $T$ can be written as
\begin{equation}
P(t_{\rm m}|T)=r F_{\rm R}\left[rt_{\rm m},r(T-t_{\rm m})\right]\,,
\label{Pres}
\end{equation}
where the double Laplace transform of the scaling function $F_{\rm R}(t_1,t_2)$ is given by \cite{prep}
\begin{eqnarray}\label{eq:LT_F_res}
\int_{0}^{\infty}dt_1\,\int_{0}^{\infty}dt_2\,F_{\rm R}(t_1,t_2)e^{-s_1 \,t_1-s_2 \,t_2} &=&\frac{1}{2} \frac{1}{\sqrt{1+s_2}\left(1+\sqrt{1+s_1}\right)}\\
&+&\frac12\frac{ \sqrt{1+s_2}}{\sqrt{s_1+1}-1}\int_{0}^{\infty}dz\,\frac{e^{-(1+\sqrt{1+s_1})z}\left(s_1 e^{\sqrt{1+s_1}z}-\sqrt{s_1+1}+1 \right)}{\left(s_1+e^{-\sqrt{1+s_1}z}\right)\left(s_2+e^{-\sqrt{1+s_2}z}\right)}\,.\nonumber
\end{eqnarray}
From this result in Eq. \eqref{eq:LT_F_res} it is easy to check that $P(t_{\rm m}|T)$ is correctly normalized to unity. Moreover, since the expression on the right-hand side of Eq. \eqref{eq:LT_F_res} is not invariant under exchange of $s_1$ and $s_2$, we find that $F_{\rm R}(t_1,t_2)\neq F_{\rm R}(t_2,t_1)$ and thus that $P(t_{\rm m}|T)$ is not symmetric around the midpoint $t_{\rm m}=T/2$.

The asymmetry of the PDF $P(t_{\rm m}|T)$ in the case of BM with stochastic resetting is also confirmed by numerical simulations, 
as shown in Fig. 2b) of the main text. However, in some cases, it might be difficult to determine whether a distribution $P(t_{\rm m}|T)$, obtained from simulations or experiments, is symmetric or not, due to measurement or statistical noise. One simpler quantity that one can study is the average value of $t_{\rm m}$. Indeed, finding that $\langle t_{\rm m}\rangle\neq T/2$ is sufficient to conclude that the full distribution $P(t_{\rm m}|T)$ is not symmetric around $t_{\rm m}=T/2$ and thus that the process is nonequilibirum. In the case of resetting BM, using Eq. \eqref{Pres}, we find that the value of $\langle t_{\rm m}\rangle(T)$ as a function of $T$ is given by \cite{prep}
\begin{equation}
\langle t_{\rm m}\rangle(T)=\frac{1}{r}f(rT)\,,
\label{avg}
\end{equation}
where the scaling function $f(t)$ is given by
\begin{eqnarray} \label{foft}
f(t)&=&\frac{1}{96}\left[-4t(2t^2+3t-18)+\frac{2}{\sqrt{\pi}}\sqrt{t}(3+16t+4t^2)e^{-t}+(-3-30t+36t^2+8t^3)\operatorname{erf}(\sqrt{t})\right]\nonumber\\ &+&\frac{1}{2}\left[e^{-t}-\frac{2}{\sqrt{\pi}}\Gamma\left(\frac32,t\right)\right]+\sum_{k=1}^{\infty} g_k(t)\,,
\end{eqnarray}
where $\Gamma(a,t) = \int_t^\infty x^{a-1}\, e^{-x}\, dx$ is the upper incomplete Gamma function,
\begin{equation}
g_k(t)=(-1)^k\frac12 (k+1)(k+2)\int_{0}^{t}d\tau\,h_k(t-\tau)\tau^{k+1}\left(\frac{1}{(k+1)!}+\frac{\tau}{(k+2)!}\right)\,,
\end{equation}
and
\begin{eqnarray}
h_k(t)&=&\frac{1}{k^2}\left\{-e^{-t+t/k^2}k(1-k)^2+e^{-t}\frac{k\left[k(1+k)^3-2k^3 t\right]}{\sqrt{\pi t}(1+k)^3}\right\}+\frac{1}{k^2}\left[\operatorname{erf}\left(\frac{\sqrt{t}}{k}\right)e^{-t+t/k^2}(1-k)^2\right]\nonumber \\
&\times &\frac{1}{(1+k)^4}e^{-t+t/(1+k)^2}\left[(1+k)^2 (k^2-2)+2kt\right]\left[1-\operatorname{erf}\left(\frac{\sqrt{t}}{(k+1)}\right)\right]\,.
\end{eqnarray}

\begin{figure}[t]
\includegraphics[scale=0.5]{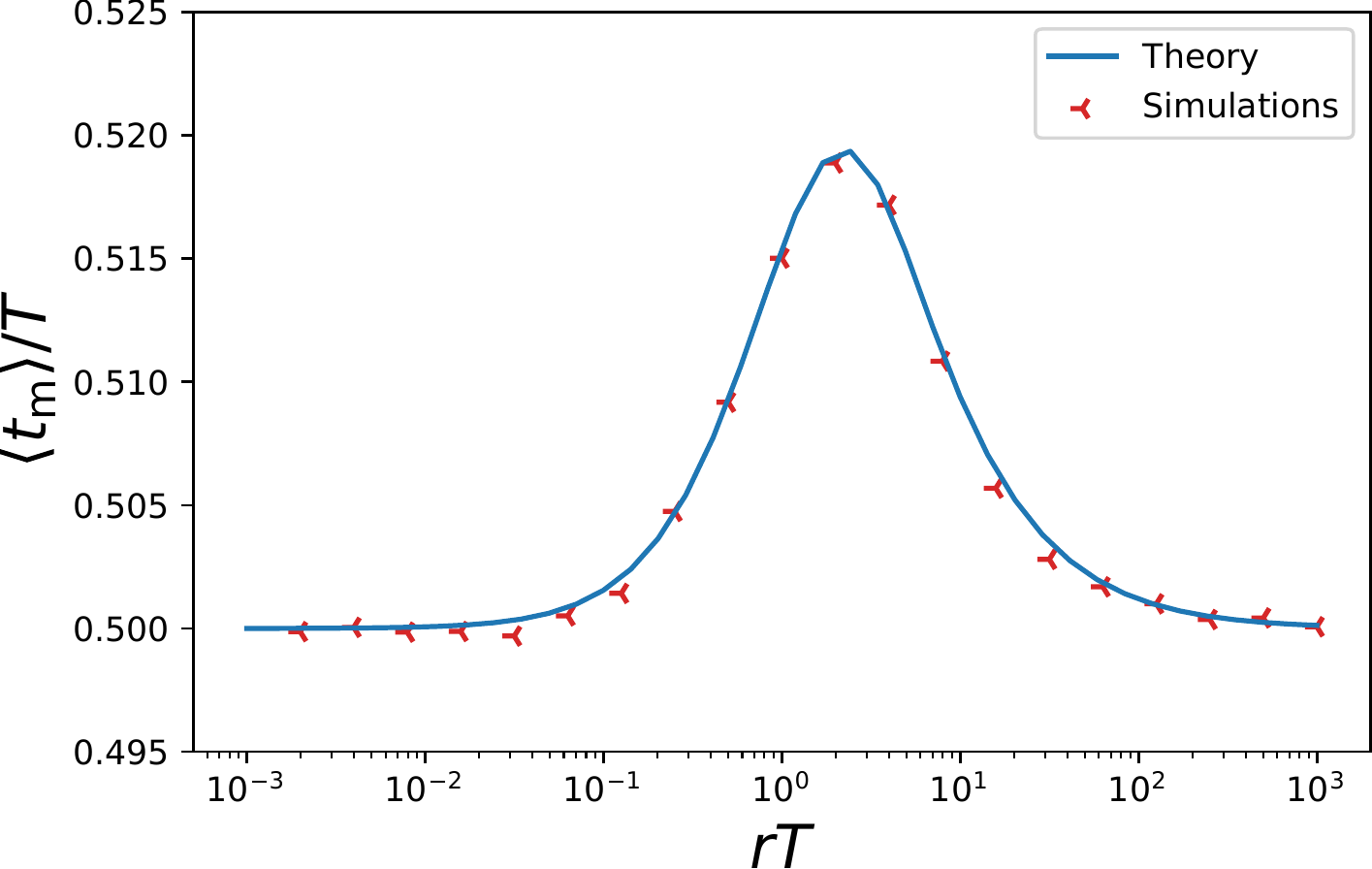} 
\caption{\label{fig:avg_tmax} The scaled average $\langle t_{\rm m}\rangle/T$ as a function of $rT$ for Brownian motion with resetting rate $r$. The symbols depict the results of numerical simulations while the continuous line correspond to the analytical result in Eq. \eqref{avg}. In the case of an equilibrium process, one expects $\langle  t_{\rm m}\rangle/T=1/2$ for any $T$.}
\end{figure}

The exact result in Eqs. \eqref{avg} and \eqref{foft} is shown in Fig. \ref{fig:avg_tmax} and is in good agreement with numerical simulations. Note that in the case of an equilibrium process one has $\langle t_{\rm m}\rangle (T)/T=1/2$. In Fig. \eqref{fig:avg_tmax}, we observe that the ratio $\langle t_{\rm m}\rangle/T$ is manifestly different from the constant value $1/2$, signaling the nonequilibrium nature of the resetting process. Note also that the deviation of $\langle t_{\rm m}\rangle/T$ from the equilibrium value $1/2$ has a maximum at some finite value of $rT$. Thus, keeping $T$ fixed, there exists an optimal value of the resetting rate $r$ that maximizes the deviation from the equilibrium result.

\subsection{Run-and-tumble particle in a confining potential}

We next consider a single RTP moving in a one-dimensional potential $V(x)=\alpha |x|^p$. The position $x(\tau)$ of the particle evolves according to the stochastic differential equation
\begin{equation}
\frac{dx(\tau)}{d\tau}=-V'(x)+v_0 \sigma(\tau)\,,
\end{equation}
where $v_0>0$ is the speed of the particle and $\sigma(\tau)=\pm 1$ is telegraphic noise, switching sign with constant rate $\gamma$. In reference \cite{DKM19} it has been shown that the nonequilibrium steady state of the system depends on the system parameters $\alpha$ and $p$. In this Section, we focus on the special case $p=1$ and $v_0>\alpha$. In this case, the steady-state probability $P_{\rm st}^{\pm}(x)$ that the particle is at position $x$ with velocity $\pm v_0$ can be written as \cite{DKM19} 
\begin{equation}
P_{\rm st}^{\pm }(x)=\frac12 \left(1\pm \frac{\alpha}{v_0}\operatorname{sign}(x)\right)\frac{\gamma~\alpha}{v_0^2-\alpha^2}\exp\left(-\frac{2\gamma\alpha}{v_0^2-\alpha^2}|x|\right)\,.
\label{eq:stationary_RTP_x_s}
\end{equation}
Note that $\int_{-\infty}^\infty P_{\rm st}^{\pm }(x) \;, dx  =1/2$. Therefore, in the stationary state, the right movers (the particles with positive velocity $+v_0$) and the left movers (the particles with positive velocity $-v_0$) occur with equal probability $1/2$. We assume that at the initial time the position $x_0$ of the particle and its velocity $v_0 \sigma(0)$ are jointly drawn from the steady state \eqref{eq:stationary_RTP_x_s}. We find that the distribution of the time $t_{\rm m }$ at which the position of the particle reaches its maximal value up to time $T$ can be written as \cite{prep}
\begin{equation}
P(t_{\rm m}|T)=P_0(T)\delta(t_{\rm m})+P_{\rm bulk}(t_{\rm m}|T)+P_1(T)\delta(t_{\rm m}-T)\,.
\label{ptm}
\end{equation}
In other words, with finite probability $P_0(T)$ the maximal position occurs at the initial time and similarly with probability $P_1(T)$ the time of the maximum will be the final time $T$. These two delta functions in the distribution of $t_{\rm m }$ are a direct consequence of the persistent nature of the RTP motion. The Laplace transforms with respect to $T$ of the amplitudes $P_0(T)$ and $P_1(T)$ are given by  
\begin{equation}
\int_{0}^{\infty} dT~P_0(T)e^{-sT}=\int_{-\infty}^{\infty}dx_0~P_{\rm st}^{-}(x_0)\tilde{Q}(x_0,s)\,,
\end{equation}
and
\begin{equation}
\int_{0}^{\infty} dT~P_1(T)e^{-sT}=\sum_{\sigma=\pm}\int_{-\infty}^{\infty}dx_0~P_{\rm st}^{\sigma}(x_0)~\tilde{G}(M,s|x_0,\sigma)\,,
\end{equation}
where $P_{\rm st}^{\sigma}(x_0)$ is given in Eq. \eqref{eq:stationary_RTP_x_s}. Here, $\tilde{Q}(x_0,s)$ and $\tilde{G}(M,s|x_0,\pm)$ denote the Laplace transforms with respect to $T$ of $Q(x_0,T)$ and $G(M,T|x_0,\pm)$, respectively. The function $Q(x_0,T)$ is defined as the probability that the RTP, starting from position $x_0$ with negative velocity, remains below its starting position $x_0$ up to time $T$. Similarly, $G(M,T|x_0,\pm)$ is the probability that the RTP, starting from position $x_0$ with velocity $\pm v_0$, reaches position $M>x_0$ for the first time at time $T$. The exact expressions of $\tilde{Q}(x_0,s)$ and $\tilde{G}(M,s|x_0,\pm)$ are given below.

The PDF $P_{\rm bulk}(t_{\rm m}|T)$ in Eq. \eqref{ptm} describes the probability density of $t_{\rm m}$ when $0<t_{\rm m}<T$. Its double Laplace transform with respect to $t_1=t_{\rm m}$ and $t_2=T-t_{\rm m}$ can be written as
\begin{equation}
\int_{0}^{\infty}dt_1~\int_{0}^{\infty}dt_2~P_{\rm bulk}(t_{\rm m}=t_1|T=t_1+t_2)e^{-s_1t_1-s_2t_2}=~\gamma~\sum_{\sigma=\pm }\int_{-\infty}^{\infty}dx_0~P_{\rm st}^{\sigma}(x_0)~\int_{x_0}^{\infty}dM~\tilde{G}(M,s_1|x_0,\sigma)~\tilde{Q}(M,s_2)\,,
\label{eq:Pbulk}
\end{equation}
where $P_{\rm st}^{\sigma}(x_0)$ is given in Eq. \eqref{eq:stationary_RTP_x_s},
\begin{equation}
{\tilde G}(M,s|x_0,+)=\begin{cases}
\displaystyle  \frac{1}{v_0+\alpha}e^{-(k-(s+\gamma)\alpha)(M-x_0)/(v_0^2-\alpha^2)}&\;{\rm for}\;x_0<0~,M<0\\
\\
\displaystyle k \frac{e^{-(\alpha(s+\gamma)+k)M/(v_0^2-\alpha^2)}~e^{(-\alpha(s+\gamma)+k)x_0/(v_0^2-\alpha^2)}}{v_0(k-\alpha(\gamma+s))+\alpha(v_0(\gamma+s)-k)e^{-2kM/(v_0^2-\alpha^2)}}&\;{\rm for}\;x_0<0~,M>0\\
\\
\displaystyle \frac{1}{v_0-\alpha} ~\frac{(k-v_0(s+\gamma))\alpha+e^{2kx_0/(v_0^2-\alpha^2)}v_0((s+\gamma)\alpha-k)}{(k-v_0(s+\gamma))\alpha+e^{2kM/(v_0^2-\alpha^2)}v_0((s+\gamma)\alpha-k)}\\ \times e^{(k-\alpha(s+\gamma))(M-x_0)/(v_0^2-\alpha^2)}&\;{\rm for}\;x_0>0~,M>0\,,\\
\\
\end{cases}
\label{G_RTP_A}
\end{equation}
\begin{equation}
{\tilde G}(M,s|x_0,-)=\begin{cases}
\displaystyle  \frac{v_0(\gamma+s)-k}{\gamma(v_0^2-\alpha^2)}e^{-(k-(s+\gamma)\alpha)(M-x_0)/(v_0^2-\alpha^2)}&\;{\rm for}\;x_0<0~,M<0\\
\\
\displaystyle \frac{k(v_0(\gamma+s)-k)}{\gamma(v_0-\alpha)} \frac{e^{-(\alpha(s+\gamma)+k)M/(v_0^2-\alpha^2)}~e^{(-\alpha(s+\gamma)+k)x_0/(v_0^2-\alpha^2)}}{v_0(k-\alpha(\gamma+s))+\alpha(v_0(\gamma+s)-k)e^{-2kM/(v_0^2-\alpha^2)}}&\;{\rm for}\;x_0<0~,M>0\\
\\
\displaystyle \frac{v_0(s+\gamma)-k}{\gamma(v_0^2-\alpha^2)} ~\frac{(k-v_0(s+\gamma))\alpha+e^{2kx_0/(v_0^2-\alpha^2)}v_0((s+\gamma)\alpha-k)}{(k-v_0(s+\gamma))\alpha+e^{2kM/(v_0^2-\alpha^2)}v_0((s+\gamma)\alpha-k)}\\ \times e^{(k-\alpha(s+\gamma))(M-x_0)/(v_0^2-\alpha^2)}&\;{\rm for}\;x_0>0~,M>0\,,\\
\\
\end{cases}
\label{G_RTP_B}
\end{equation}
and
\begin{equation}
{\tilde Q}(M,s)=\begin{cases}
\displaystyle  \frac{1}{s}\frac{k+v_0s-\gamma\alpha}{k+v_0(s+\gamma)}&\;{\rm for}\;M<0\\
\\
\displaystyle \frac{1}{s}\frac{1}{k+v_0(s+\gamma)}\left[k+v_0s+\alpha\gamma-\frac{2k\gamma\alpha(v_0-\alpha)}{(v_0(s+\gamma)-k)\alpha+v_0(k-(s+\gamma)\alpha)e^{2kM/(v_0^2-\alpha^2)}}\right]&\;{\rm for}\;M>0\,.\\
\\
\end{cases}
\label{Q_RTP}
\end{equation}
In the equations above we have defined
\begin{equation}
k=\sqrt{s^2v_0^2+2sv_0^2\gamma+\gamma^2\alpha^2}\,.
\label{eq:k}
\end{equation}

From the equations above it is possible to show (e.g. by numerical integration) that $P(t_{\rm m}|T)$ in Eq. \eqref{ptm} is correctly normalized to unity. Moreover, for $0<t_{\rm m}<T$ one can check that $P(t_{\rm m}|T)=P(T-t_{\rm m}|T)$, i.e., that the central part of the distribution of $t_{\rm m}$ is symmetric around the midpoint $t_{\rm m}=T/2$. However, it is easy to show that the amplitudes $P_0(T)$ and $P_1(T)$ of the delta functions in $t_{\rm m}=0$ and $t_{\rm m}=T$ are not equal. Thus, the full distribution $P(t_{\rm m}|T)$, for $0\leq t_{\rm m}\leq T$ is not symmetric around $t_{\rm m}=T/2$. This is in agreement with the criterion presented in the main text, since the process is out-of-equilibrium.

\section{Maximum of a confined Brownian particle at late times}

We consider a Brownian particle in a potential that grows as $V(x) \simeq \alpha |x|^p$ for large $|x|$, with $p>0$. At the initial time we assume that the particle starts from position $x_0$ and evolves according to the Langevin equation \eqref{sup:langevin} up to time $T$. We assume that the initial position $x_0$ is drawn from the equilibrium steady state
\begin{equation}
P_{\rm st}(x_0)\propto \exp\left(-\frac{V(x_0)}{D}\right)\,.
\label{equilibrium_app}
\end{equation}
We want to investigate the distribution of the maximal position $M$ reached by the particle up to time $T$. 

In order to estimate the distribution of $M$, we will apply the following heuristic argument. For $p\geq 1$, we expect the autocorrelation function to decay as
\begin{equation}
\langle x(\tau_1)x(\tau_2)\rangle-\langle x(\tau_1)\rangle \langle x(\tau_2)\rangle\sim e^{-|\tau_1-\tau_2|/T_B}\,.
\end{equation}
We thus divide the time interval $[0,T]$ in $N_B=T/T_B$ intervals of size $T_B$ and we denote by $m_i$ the maximal position reached in the $i$-th interval. Since the size of the blocks is the correlation time $T_B$, the variables $m_1\,,\ldots\,,m_N$ can be considered independent. The global maximum $M$ is given by
\begin{equation}
M=\max_{1\leq i\leq N}\left[m_1\,,m_2\,,\ldots\,,m_N\right]\,.
\end{equation}
Even if we do not know the PDF $P(m)$ of the local maxima $m_i$, we can guess that it will have the same right tail as the equilibrium distribution in Eq. \eqref{equilibrium_app}, i.e., that for large $m$ one has
\begin{equation}
P(m)\sim \exp\left(-\frac{\alpha }{D}~ m^p\right)\,.
\end{equation}
Thus, one can apply the standard extreme value theory for i.i.d. random variables (see, e.g., Ref. \cite{MP20}) and one finds that, for large $T$
\begin{equation}
M = \left[\frac{D}{\alpha}\log(T)\right]^{1/p}+O(1)\,.
\end{equation}
In other words, for $T\gg T_B$, the maximum of the process becomes to leading order deterministic, with subleading random fluctuations of order one.

\section{Time of the maximum for Gaussian stationary processes}

In this section, we show that for any Gaussian stationary process, the distribution $P(t_{\rm m}|T)$ of the time $t_{\rm m}$ of the maximum is symmetric around its midpoint $t_{\rm m}=T/2$, i.e., that $P(t_{\rm m}|T)=P(T-t_{\rm}|T)$. Let us consider a one-dimensional discrete-time Gaussian stationary process $x_k$ with $1\leq k \leq T$. Note that here we assume that $T$ is an integer number. The derivation below can be easily generalized to continuous-time processes. For simplicity we assume that the average value of the process is zero, i.e., that $\langle x_k\rangle=0$ for any $k$. The probability of observing a given trajectory $\{x_k\}=\{x_1\,,\ldots\,,x_T\}$ is given by
\begin{equation}
P(\{x_k\})=N ~\exp\left[-\frac12 \sum_{i,j} ~x_i \Sigma^{-1}_{i,j}x_j\right]\,,
\label{Pxk_1}
\end{equation}
where $\Sigma_{i,j}=\langle x_i x_j\rangle$ is the covariance matrix and $N$ is a normalization constant. By definition of Gaussian stationary process, the covariance $\Sigma_{i,j}$ only depends on $|i-j|$, thus the expression in Eq. \eqref{Pxk_1} can be rewritten as
\begin{equation}
P(\{x_k\})=N ~\exp\left[-\frac12 \sum_{i,j} ~x_i \Sigma^{-1}(|i-j|)x_j\right]\,.
\label{Pxk_2}
\end{equation}
Let us now consider the time-reversed trajectory $\{\bar{x}_k\}=\{x_{T-k}\}$. The probability of observing the trajectory $\{\bar{x}_k\}$ is given by
\begin{equation}
P(\{\bar{x}_k\})=N ~\exp\left[-\frac12 \sum_{i,j} ~x_{T-i} \Sigma^{-1}(|i-j|)x_{T-j}\right]\,.
\label{Pxk_3}
\end{equation}
Performing the change of variable $(i,j)\to(i'=T-i,j'=T-j)$, we obtain
\begin{equation}
P(\{\bar{x}_k\})=N ~\exp\left[-\frac12 \sum_{i',j'} ~x_{i'} \Sigma^{-1}(|i'-j'|)x_{j'}\right]\,.
\label{Pxk_4}
\end{equation}
Comparing this expression with Eq. \eqref{Pxk_2}, we finally get
\begin{equation}
P(\{\bar{x}_k\})=P(\{x_k\})\,.
\end{equation}
In other words, the process is symmetric under time reversal. As we have shown in the main text, this implies that the distribution of $t_{\rm m}$ is symmetric around $t_{\rm m}=T/2$, i.e., that $P(t_{\rm m}|T)=P(T-t_{\rm m}|T)$.

As an example of Gaussian stationary process, we consider a single active Ornstein-Uhlenbeck particle (AOUP) in a one dimensional harmonic potential $V(x)=\alpha x^2$ \cite{FNC16}. The position $x(\tau)$ of the AUOP evolves according to
\begin{equation}
\frac{dx(\tau)}{d\tau}=-\alpha x(\tau)+v(\tau)+\sqrt{2D}\xi(\tau)\,,
\label{AOUP_x}
\end{equation}
where $\xi(t)$ is a Gaussian white noise with zero mean and correlator $\langle \xi(\tau)\xi(\tau')\rangle=\delta(\tau-\tau')$ and the active noise $v(\tau)$ is a Ornstein-Uhlenbeck process. In other words, $v(\tau)$ evolves according to
\begin{equation}
\frac{dv(\tau)}{d\tau}=-\frac{v}{\tau_a}+\frac{\sqrt{2 D_a}}{\tau_a}\zeta(\tau)\,,
\label{AOUP_v}
\end{equation}
where $D_a$ and $\tau_a$ are positive constants and $\zeta(\tau)$ is a Gaussian white noise, uncorrelated with $\xi(\tau)$. Note that $\tau_a>0$ can be interpreted as the persistence time of the driving noise $v(\tau)$, which induces memory in the evolution of the position $x(\tau)$.

Note that while the evolution of $x(\tau)$ is influenced by $v(\tau)$, there is no feedback mechanism from $x(\tau)$ to $v(\tau)$. This induces a net probability current in the phase space $(x,v)$, which violates detailed balance. Thus, the system is out-of-equilibrium \cite{DBE20}. However, since Eqs. \eqref{AOUP_x} and \eqref{AOUP_v} are linear, it is clear that $x(\tau)$ is a Gaussian stationary process (if initialized from the stationary state). Thus, from the discussion above the distribution $P(t_{\rm m}|T)$ of the time $t_{\rm m}$ of the maximum of the position $x(\tau)$ is symmetric around $t_{\rm m}=T/2$, even if the joint process $(x,v)$ is out-of-equilibrium.

\end{widetext}

\end{document}